\documentclass[traditabstract,longauth]{aa}

\newcommand{\ud}{\ensuremath{\mathrm{d}}\xspace}
\newcommand{\pvec}{\ensuremath{\vec{\theta}}\xspace}
\newcommand{\data}{\ensuremath{\vec{D}}\xspace}
\newcommand{\covmat}{\ensuremath{\bf{\Sigma}}\xspace}

\usepackage{xspace}

\usepackage{graphicx}

\usepackage{txfonts}

\usepackage{natbib}
\bibpunct{(}{)}{;}{a}{}{,} 
\defcitealias{leger2009}{L{\'e}ger, Rouan, Schneider et~al. 2009}

\def\ms{\hbox{\,m\,s$^{-1}$}}         
\def\m2s2{\hbox{\,m$^{2}$\,s$^{-2}$}} 
\def\kms{\hbox{\,km\,s$^{-1}$}}       
\def\gcm3{\hbox{\,g\,cm$^{-3}$}}      

\begin{document}

\title{Transiting exoplanets from the CoRoT space mission\thanks{The CoRoT
  space mission, launched on December 27th 2006, has
  been developed and is operated by CNES, with the contribution of
  Austria, Belgium, Brazil, ESA (RSSD and Science Programme), Germany
  and Spain. 
  Based on observations made with HARPS ({\it {High Accuracy Radial 
  velocity Planet Searcher}}) spectrograph on the
  3.6-m European Organisation for Astronomical Research in the
  Southern Hemisphere telescope at La Silla Observatory, Chile (ESO
  program 188.C-0779).
  },\thanks{Based on observations obtained with the Nordic Optical Telescope, 
operated on the island of La Palma jointly by Denmark, Finland, Iceland, 
Norway, and Sweden, in the Spanish Observatorio del Roque de los Muchachos 
of the Instituto de Astrofisica de Canarias, in time allocated by 
the Spanish Time Allocation Committee (CAT).
}
}

\subtitle{XXVIII. CoRoT-33b, an object in the brown dwarf desert with 2:3 commensurability 
with its host star}


\author{Sz.~Csizmadia\inst{\ref{DLR}}
\and A.~Hatzes\inst{\ref{Tautenburg}} 
\and D.~Gandolfi\inst{\ref{Torino}, \ref{LSW}}
\and M.~Deleuil\inst{\ref{LAM}} 
\and F.~Bouchy\inst{\ref{LAM}} 
\and M.~Fridlund\inst{\ref{MPIA},\ref{Leiden},\ref{Onsala}}
\and L.~Szabados\inst{\ref{Konkoly}}
\and H.~Parviainen\inst{\ref{OxfordH}}
\and J.~Cabrera\inst{\ref{DLR}} 
\and S.~Aigrain\inst{\ref{Oxford}} 
\and R.~Alonso\inst{\ref{IAC},\ref{Laguna}} 
\and J.-M.~Almenara\inst{\ref{LAM}} 
\and A.~Baglin\inst{\ref{LESIA}}
\and P.~Bord\'e\inst{\ref{LAB}} 
\and A.~S.~Bonomo\inst{\ref{Torino2}}  
\and H.~J.~Deeg\inst{\ref{IAC},\ref{Laguna}} 
\and R.~F.~D{\'i}az\inst{\ref{Geneva}}
\and A.~Erikson\inst{\ref{DLR}}
\and S.~Ferraz-Mello\inst{\ref{Brasil}} 
\and M.~Tadeu~dos~Santos\inst{\ref{Brasil}} 
\and E.~W.~Guenther\inst{\ref{Tautenburg}} 
\and T.~Guillot\inst{\ref{OCA}} 
\and S.~Grziwa\inst{\ref{Koeln}} 
\and G.~H\'ebrard\inst{\ref{IAP}} 
\and P.~Klagyivik\inst{\ref{IAC},\ref{Laguna}} 
\and M.~Ollivier\inst{\ref{IAS}} 
\and M.~P\"atzold\inst{\ref{Koeln}} 
\and H.~Rauer\inst{\ref{DLR},\ref{ZAA}} 
\and D.~Rouan\inst{\ref{LESIA}}
\and A.~Santerne\inst{\ref{Porto}} 
\and J.~Schneider\inst{\ref{LUTh}} 
\and T.~Mazeh\inst{\ref{Telaviv}}
\and G.~Wuchterl\inst{\ref{Tautenburg}}
\and S.~Carpano\inst{\ref{Heidelberg}}
\and A. Ofir\inst{\ref{Weizmann}}
}

%

\institute{
Institute of Planetary Research, German Aerospace Center, Rutherfordstrasse 2, 12489 Berlin, Germany\label{DLR}
\and Th\"uringer Landessternwarte, Sternwarte 5, Tautenburg 5, D-07778 Tautenburg, Germany\label{Tautenburg}
\and Dipartimento di Fisica, Universit\'a di Torino, via P. Giuria 1, 10125 Torino, Italy\label{Torino}
\and Landessternwarte K\"onigstuhl, Zentrum f\"ur Astronomie der Universit\"at Heidelberg, K\"onigstuhl 12, D-69117 Heidelberg, Germany\label{LSW}
\and Aix Marseille Universit\'e, CNRS, LAM (Laboratoire d'Astrophysique de Marseille) UMR 7326, 13388, Marseille, France\label{LAM}
\and Max-Planck-Institut f\"ur Astronomie, K\"onigstuhl 17, 69117 Heidelberg, Germany\label{MPIA}
\and Leiden Observatory, University of Leiden, PO Box 9513, 2300 RA, Leiden, The Netherlands\label{Leiden}
\and Department of Earth and Space Sciences, Chalmers University of Technology, Onsala Space Observatory, 439 92, Onsala, Sweden\label{Onsala}
\and Konkoly Observatory of the Hungarian Academy of Sciences, MTA CSFK, Budapest, Konkoly Thege Mikl\'os \'ut 15-17, Hungary\label{Konkoly}
\and Sub-department of Astrophysics, Department of Physics, University of Oxford, Oxford, OX1 3RH, UK\label{OxfordH}
\and Department of Physics, Denys Wilkinson Building, Keble Road, Oxford, OX1 3RH\label{Oxford}
\and Instituto de Astrof{\'i}sica de Canarias, E-38205 La Laguna, Tenerife, Spain\label{IAC}
\and Universidad de La Laguna, Dept. de Astrof\'isica, E-38206 La Laguna, Tenerife, Spain\label{Laguna}
\and LESIA, UMR 8109 CNRS, Observatoire de Paris, UPMC, Universit\'e Paris-Diderot, 5 place J. Janssen, 92195 Meudon, France\label{LESIA}
\and LAB, UMR 5804, Univ. Bordeaux \& CNRS, F-33270, Floirac, France\label{LAB}
\and INAF - Osservatorio Astrofisico di Torino Strada Osservatorio, 20 10025 Pino Torinese (TO), Italy\label{Torino2}
\and Observatoire astronomique de l'Universit\'e de Gen\`eve, 51 ch. des Maillettes, CH-1290 Versoix, Switzerland\label{Geneva}
\and IAG-Universidade de S\~ao Paulo, Brasil\label{Brasil}
\and Universit\'e de Nice-Sophia Antipolis, CNRS UMR 6202, Observatoire de la C\^ote d'Azur, BP 4229, 06304 Nice Cedex 4, France\label{OCA}
\and Rheinisches Institut f\"ur Umweltforschung an der Universit\"at zu K\"oln, Aachener Strasse 209, 50931, Germany\label{Koeln}
\and Institut d'Astrophysique de Paris, UMR 7095 CNRS, Universit\'e Pierre \& Marie Curie, 98bis boulevard Arago, 75014 Paris, France\label{IAP}
\and Institut d'Astrophysique Spatiale, Universit\'e Paris XI, F-91405 Orsay, France\label{IAS}
\and Center for Astronomy and Astrophysics, TU Berlin, Hardenbergstr. 36, 10623 Berlin, Germany\label{ZAA}
\and Instituto de Astrof\'isica e Ci\^{e}ncias do Espa\c co, Universidade do Porto, CAUP, Rua das Estrelas, P-4150-762 Porto, Portugal\label{Porto}
\and LUTH, Observatoire de Paris, UMR 8102 CNRS, Universit\'e Paris Diderot; 5 place Jules Janssen, 92195 Meudon, France\label{LUTh}
\and School of Physics and Astronomy, Raymond and Beverly Sackler Faculty of Exact Sciences, Tel Aviv University, Tel Aviv, Israel\label{Telaviv}
\and Max-Planck-Institut f\"ur extraterrestrische Physik, Giessenbachstrasse 1, 85748 Garching, Germany\label{Heidelberg}
\and Department of Earth and Planetary Sciences, Weizmann Institute of Science, 234 Herzl St., Rehovot 76100, Israel\label{Weizmann}
}
\date{Received 16 June 2015; accepted 28 July 2015}



\abstract{We report the detection of a rare transiting brown dwarf with a mass 
          of 59 $M_\mathrm{Jup}$ and radius of 1.1 $R_\mathrm{Jup}$ around the metal-rich, ${\rm [Fe/H]}=+0.44$, G9V star 
	  CoRoT-33. The orbit is eccentric ($e=0.07$) with a period of 5.82 d. The companion,
          CoRoT-33b, is thus a new member in the so-called brown 
	  dwarf desert. The orbital period is within 3\% to a 3:2 resonance 
	  with the rotational period of the star. CoRoT-33b may be an important test case for tidal 
	  evolution studies. The true frequency of brown dwarfs 
	  close to their host stars ($P<10$~d) is estimated to be approximately 0.2\% 
	  which is about six times smaller than the frequency of hot 
	  Jupiters in the same period range. We suspect that the frequency of brown dwarfs declines 
	  faster with decreasing period than that of giant planets.}


\titlerunning{CoRoT-33b: a synchronized brown dwarf}
\authorrunning{Csizmadia et~al.}

\maketitle

%
\section{Introduction}
\label{sec:introduction}

As of June 2015, W.~R. Johnston lists 2085 confirmed and 562 candidate brown 
dwarfs\footnote{See at 
http://www.johnstonsarchive.net/astro/browndwarflist.html). The 
DwarfArchive.org project, maintained by C.~Gelino, D.~Kirkpatrick, 
M.~Cushing, D.~Kinder, 
A.~Burgasser (http://spider.ipac.caltech.edu/staff/davy/ARCHIVE/index.shtml) 
tabulates 1281 L, T and Y dwarfs but it does not contain information 
about the masses and radii of the catalogued objects. The 
Johnston's catalogue reports radius values for 191 brown dwarfs, but most 
of these were obtained by several different and often indirect methods, 
e.g. model-fitting to the spectral energy distributions. Radii of brown 
dwarfs measured by direct methods are listed in Table 1.}, 427 of them 
are marked as located in either a binary or multiple system. Only 65 brown dwarfs 
are companions to FGK dwarfs at orbital distances less than 2 AU from 
the star (Ma \& Ge 2014). Of these only nine (and one additional reported here) are transiting  
 (Table~1) which provides the possibility of measuring directly the mass and 
radius directly. Given the meager 
observational data, an increase in the number of brown dwarfs with 
accurate parameters is necessary to obtain a better 
understanding of the formation, structure, and evolution of these 
objects. Transiting brown dwarfs offer the best possibility for  
characterizing them in a model-independent way.

The radii of brown dwarfs and giant planets are in the same range
around 1 $R_{Jup}$, but the masses are quite different. Brown dwarfs 
straddle a border at 65 Jupiter masses. Objects with masses 
below this border fuse deuterium (D) while those above this mass limit fuse 
lithium (Li) in episodic events (e.g. Chabrier et~al. 1996), and planets 
do not ignite their material. Spiegel 
et al. (2011) showed that the lower mass limit for brown dwarfs 
-- defined by deuterium-ignition -- is between 11-16 $M_\mathrm{Jup}$ 
depending on the actual metallicity. Planets are usually smaller in mass 
than brown dwarfs, but there is an overlapping region: a tiny portion of 
planetesimals, the so-called `super-planets' may grow up to 20-40 
$M_\mathrm{Jup}$ by core-accretion (Mordasini et al. 2009), 
causing problems with separating these giant planets from brown dwarfs 
(Schneider et al. 2011). The maximum mass of brown dwarfs is about 
75-80 $M_\mathrm{Jup}$ (Baraffe et al. 2002).

Chabrier et~al. (2014) suggests that the borderline between giant planets and 
brown dwarfs should be linked to their different formation scenarios and not 
to the minimum mass required for deuterium ignition. Then the various formation 
and evolutionary mechanisms may be responsible for the nonequal frequency of brown 
dwarfs and planetary-mass objects around stars.

While $\sim$50\% of solar-like dwarf stars have a stellar, $\sim$30\% 
have a low-mass (Earth- to Neptune-sized), and $\sim$2.5\% have a higher 
mass (Jupiter or bigger) planetary companion, only 0.6-0.8\% of 
solar-like stars have a brown dwarf within 5~AU (Duquennoy \& Mayor 
1991; Vogt et~al. 2002; Patel et~al. 2007; Wittenmyer et~al. 2009; 
Sahlmann et~al. 2011; Dong \& Zhu 2013). The brown dwarf desert refers 
to this low occurrence rate of brown dwarfs as companion objects to 
main-sequence stars. However, the frequency of wide pairs consisting of a 
brown dwarf and a solar-like star (with a separation exceeding 5~AU) is 
significantly higher at over 2-3\% (Ma \& Ge 2014 and references 
therein). A more detailed overview of the different occurrence rates can 
be found in Ma \& Ge (2014).

Brown dwarfs may have a different formation mechanism from giant planets, which 
might form via core accretion (Alibert et~al. 2005). Ma \& Ge (2014) proposed 
that brown dwarfs below 35 M$_\mathrm{Jup}$ are formed from the protoplanetary 
disk via gravitational instability, while those with $M$ $>$ 55 M$_\mathrm{Jup}$ 
form like stellar binaries via molecular cloud fragmentation. Between masses
of 35 and 55 M$_\mathrm{Jup}$~ there are a significant lack of brown 
dwarfs orbital periods shorter than 100 days (Ma \& Ge 2014) as 
companions to stars. 

Alternative theories explaining the brown dwarf 
desert suggest that during the formation of a binary system 
consisting of a solar-type star and a brown dwarf, the migration 
process is so effective that brown dwarf companion spirals into the 
star and is engulfed. This would explain the paucity of very few 
brown dwarf companions found within 5~AU to solar-type stars 
(Armitage \& Bonnell 2002). Combining this low frequency of close-in 
brown dwarfs with the geometric transit probability, transiting 
brown dwarfs should indeed be rare.

In this paper we report the discovery of a transiting brown dwarf,
CoRoT-33b. With a mass of 59 Jupiter mass, CoRoT-33b
lies just below to the D-Li border for brown dwarfs at 65 Jupiter masses.
We also give a preliminary estimate for the frequency of close-in
brown dwarfs with orbital period less than ten days.

\begin{table*}
\caption{Basic data of known transiting brown dwarfs. $\rho$ is the mean density of the brown dwarf component.}
\centering
\begin{tabular}{l@{\hskip1mm}c@{\hskip1mm}c@{\hskip1mm}c@{\hskip1mm}c@{\hskip1mm}c@{\hskip1mm}c@{\hskip2mm}c@{\hskip2mm}c@{\hskip2mm}c@{\hskip2mm}c@{\hskip2mm}c@{\hskip2mm}}
\hline\hline                 
Name  & $M_\mathrm{star}/M_\odot$ & $R_\mathrm{star}/R_\odot$ & $T_\mathrm{star}$ [K] & $\mathrm{[Fe/H]}$ & $P$ (days) & $e$ & $M_\mathrm{BD}/M_\mathrm{Jup}$ &$R_\mathrm{BD}/R_\mathrm{Jup}$ & $\rho$ [g/cm$^3$] & Ref. \\
\hline                                         			  
2M0535-05a$^a$  &                      &                        &                  &                   & $9.779621(42)       $ & 0.3225$\pm$0.0060           & $56.7\pm4.8$            & 6.5$\pm$0.33              & 0.26$\pm$0.06     & 1 \\
2M0535-05b$^a$  &                      &                        &                  &                   & $9.779621(42)       $ & 0.3225$\pm$0.0060           & $35.6\pm2.8$            & 5.0$\pm$0.25              & 0.35$\pm$0.08     & 1 \\
CoRoT-3b     & $1.37\pm0.09$           & $1.56\pm0.09$          & $6740\pm140$     & -0.02$\pm$0.06$^b$& $4.25680(5)         $ & 0.0                         & $21.66\pm1.0$           & $1.01\pm0.07$             & 26.4$\pm$5.6      & 2 \\
CoRoT-15b    & $1.32\pm0.12$           & $1.46^{+0.31}_{-0.14}$ & $6350\pm200$     & +0.1$\pm$0.2      & $3.06036(3)         $ & 0                           & $63.3\pm4.1$            & $1.12^{+0.30}_{-0.15}$    & 59$\pm$29         & 3 \\
CoRoT-33b    & $0.86\pm0.04$           & $0.94^{0.14}_{-0.08}$  & $5225\pm80$      & +0.44$\pm$0.10    & $5.819143(18)$        & $0.0700\pm0.0016$ 	     & $59.0^{+1.8}_{-1.7}$    & $1.10\pm0.53$             & 55$\pm27$         & 4 \\
KELT-1b      & $1.335\pm0.063$         & $1.471^{+0.045}_{-0.035}$& $6516\pm49$    & +0.052$\pm$0.079  & $1.217513(15)$        & $0.01^{+0.01}_{-0.007}$       & $27.38\pm0.93$          & $1.116^{+0.038}_{-0.029}$ & $24.5^{1.5}_{-2.1}$ & 5 \\
Kepler-39b$^c$& $1.10^{+0.07}_{-0.06}$ & $1.39^{+0.11}_{-0.10}$ & $6260\pm140$     & -0.29$\pm$0.10    & $21.0874(2)$	      & $0.121^{+0.022}_{-0.023}$     & $18.00^{+0.93}_{-0.91}$ & $1.22^{+0.12}_{-0.10}$    &12.40$^{+3.2}_{-2.6}$ & 6 \\
Kepler-39b$^c$& $1.29^{+0.06}_{-0.07}$ & 1.40$\pm$0.10        & $6350\pm100$       & +0.10$\pm$0.14     & $21.087210(37)$     & $0.112\pm0.057$               & $20.1^{+1.3}_{-1.2}$    & $1.24^{+0.09}_{-0.10}$    &13.0$^{+3.0}_{-2.2}$ & 7 \\
KOI-189b$^d$& $0.764\pm0.051$          & 0.733$\pm$0.017        & $4952\pm40$ 	   & -0.07$\pm$0.12    & $30.3604467(5)$     & $0.2746\pm0.0037$             & $78.0\pm3.4$            & $0.998\pm0.023$           & 97.3$\pm$4.1        & 8 \\
KOI-205b    & $0.925\pm0.033$	       & $0.841\pm0.020$        & $5237\pm60$      & +0.14$\pm$0.12     & $11.7201248(21)$    & $<$0.031                      & $39.9\pm1.0$            & $0.807\pm0.022$  	    & 75.6$\pm$5.2        & 9 \\
KOI-205b    & $0.96^{+0.03}_{-0.04}$   & $0.87\pm0.020$         & $5400\pm75$      & +0.18$\pm$0.12     & $11.720126(11)$     & $<$0.015                      & $40.8^{=1.1}_{-1.5}$    & $0.82\pm0.02$  	    & 90.9$^{+7.2}_{6.8}$ & 6 \\
KOI-415b    & $0.94\pm0.06$            & $1.15^{+0.15}_{-0.10}$ & $5810\pm80$      & -0.24$\pm$0.11    & $166.78805(22)$     & $0.698\pm0.002$               & $62.14\pm2.69$          & $0.79^{+0.12}_{-0.07}$    &157.4$^{+51.4}_{-52.3}$& 10 \\
LHS 6343C$^e$&$0.370\pm0.009$          & $0.378\pm0.008$        & $3130\pm20$      & +0.04$\pm$0.08    & $12.71382(4)$       & $0.056\pm0.032$               & $62.7\pm2.4$            & $0.833\pm0.021$           &109$\pm$8            & 11 \\
WASP-30b    & $1.166\pm0.026$          & $1.295\pm0.019$        & $6201\pm97$      & -0.08$\pm$0.10    & $4.156736(13)$      & 0                             & $60.96\pm0.89$          & $0.889\pm0.021$           &107.6$\pm$1.1        & 12 \\
\hline
\end{tabular}
\tablefoot{References: 
1: Stassun, Mathieu \& Valenti (2006) ; 2M0535-05
2: Deleuil et~al. (2008),  
3: Bouchy et~al. (2011a),  
4: this study,             
5: Siverd et~al. (2012),   
6: Bouchy et~al. (2011b),  
7: Bonomo et al. (2015),   
8: D\'{\i}az et~al. (2014), 
9: D\'{\i}az et~al. (2013), 
10: Moutou et~al. (2013),   
11: Johnson et~al. (2011),  
12: Anderson et~al. (2011) 
}
\tablefoot{
{\it a:} 2M0535-05 is an extreme young eclipsing system in which two brown dwarfs orbit each other.
{\it b:} $\mathrm{[M/H]}$ value is reported in the reference. Notice that $[M/H] \approxeq [Fe/H]$; we did not convert the inhomogeneous [Fe/H] to the same scale.
{\it c:} aka KOI-423b.
{\it d:} D\'{\i}az et al. (2014) concluded that KOI-189b can be either a high-mass brown dwarf or a very low mass star, too, therefore its status is uncertain.
{\it e:} the brown dwarf orbits companion A of a binary system, 
and data of the component A is given here. Star B has $M=0.30\pm0.01\mathrm{M}_\odot$, 
$T_\mathrm{eff}=3030\pm30$~K (Johnson et~al. 2011)}
\label{knownBrownDwarfs}   
\end{table*}

%
\section{Data}
\label{sec:observations}

%
\subsection{CoRoT observations}
\label{subsec:corot_observations}

\begin{figure*}[th]
\begin{center}
\resizebox{\hsize}{!}{\includegraphics[angle=0]{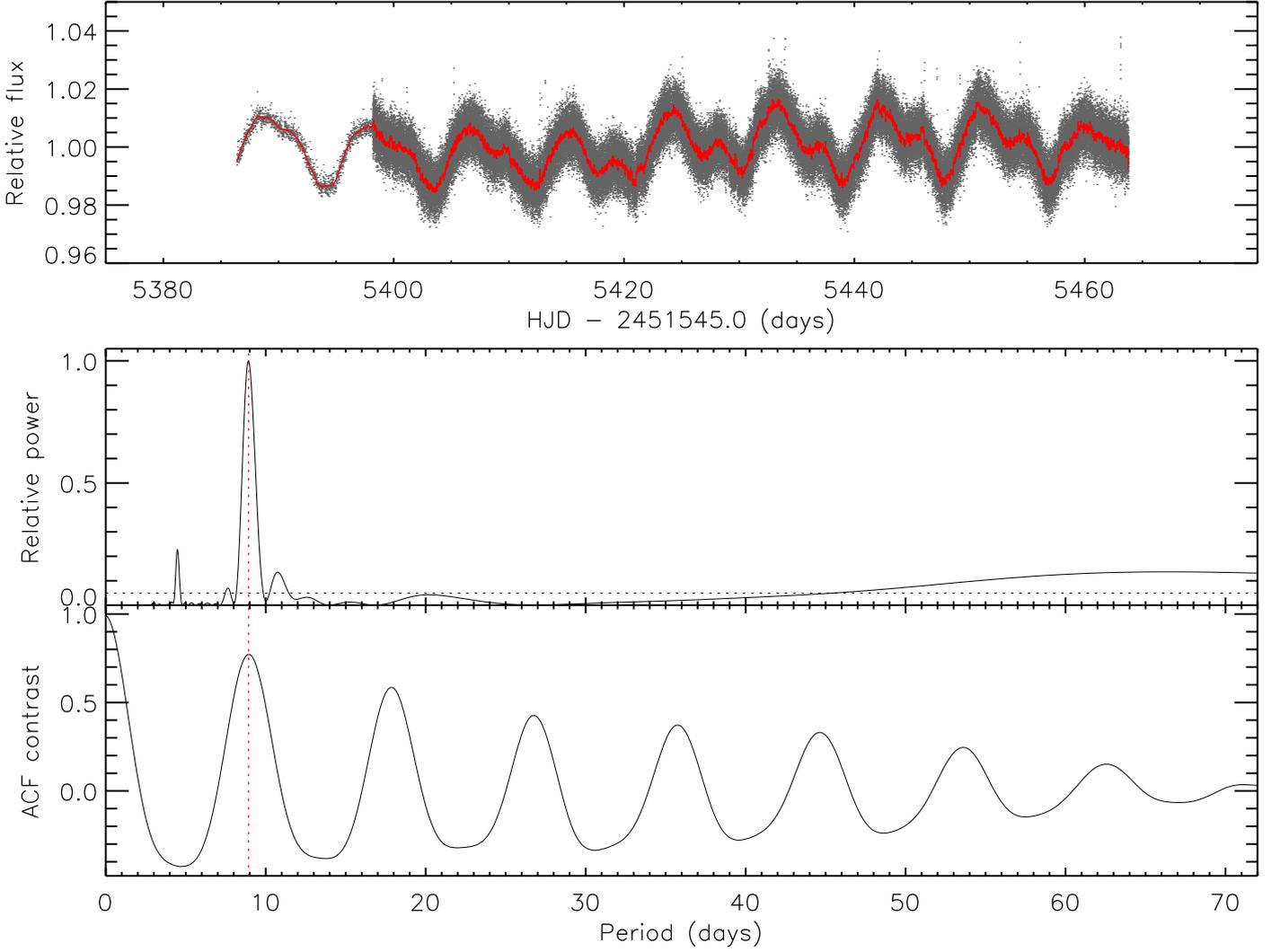}}
\caption{\emph{Upper panel:} full CoRoT light curve of CoRoT-33. The gray points represent the 
median-normalized raw data points after a 5-point width median filtering. We only used data points with 
flag ``0'' for this curve. The red line is a convolution of the raw light curve 
with a Savitzky-Golay filter that enhances the light-curve variations. \emph{Middle panel:}
Lomb-Scargle periodogram of the light curve of 
CoRoT-33. The horizontal dashed line denotes the 0.01\% false-alarm probability (Scargle 1982). The vertical 
red line marks the rotation period of the star. \emph{Lower panel:} autocorrelation function 
(ACF) of light curve, following the subtraction of the best fitting transit model. The red 
dashed line marks the peak corresponding to the rotation period of CoRoT-33 (see Section 3.5)}
\label{rawlightcurve}
\end{center}
\end{figure*}

%
\subsubsection{Flux measurements and transit detection}
\label{subsubsec:flux_measurements}

CoRoT-33 ($R=14.25$ magnitude) was observed by the CoRoT satellite 
(Auvergne et~al. 2009; Baglin et~al. 2007) in white light for 77.4 days 
between 8 July 2010 and 24 September 2010. In total, 178,342 data points 
were collected. Various designations of the target are listed in Table~2 
along with equatorial coordinates and magnitudes in different passbands.

The first transits of CoRoT-33b were discovered in the so-called Alarm 
Mode which triggered an oversampling rate and spectroscopic follow-up 
observations. After 22 July 2010 the standard 512~s sampling rate was 
changed to 32~s (see Surace et~al. 2008; Bonomo et al. 2012). In total, 
13 transits of CoRoT-33b were observed.

For the light-curve analysis we kept only those data points that 
were flagged with `$0$' by the CoRoT automatic data-pipeline which is 
an indication of good measurement without comment. In total, we used 
1626 data points obtained with the 512 s integration time and 
133,333 data points obtained with the 32 s integration time for a 
total of 134,959 photometric measurements.

The cleaned light curve normalized to
its median value is shown in Fig. 1. 
Fortunately, no significant jumps, cosmic ray events, hot pixels,
etc. affected this light curve. Also shown is a smoothed version of 
the data produced after applying a Savitzky-Golay filter. Variations 
due to rotational modulation can clearly be seen, which is  
studied in detail in Section~\ref{subsec:variability}.

%
\subsubsection{Contamination}
\label{subsubsec:contamination}

Figure~\ref{fig:contamination} shows the Palomar Observatory Sky Survey's image
of CoRoT-33 and its environment. The solid zigzag line represents the
CoRoT-photometric mask while the dashed line shows the boundaries of the
CoRoT-imagette. CoRoT has a bi-prism in the optical pathway of its
exoplanet channel so that a small, very low-resolution spectrum is obtained for
brighter stars in the field. 
Therefore the point-spread function (PSF) of each object in the
field of view of the exoplanet channel is roughly  $46\times23$ arcseconds. We
found that the contribution of stars outside the mask is very small,
but there are two main contaminants inside the photometric mask,
denoted by Nos. 1 and  2. They are fainter than CoRoT-33 by 2.5 and 2.9 magnitudes. 
Because of the large PSF of CoRoT, the contaminating sources actually produce 
13$\pm$4\% of the total observed light in this mask. This contamination factor was 
calculated using the procedure described in Pasternacki et~al. (2011) and
is consistent with the value found independently by Gardes et~al. 
(2011).

\begin{figure}
  \begin{center}
\includegraphics[width=7.62cm]{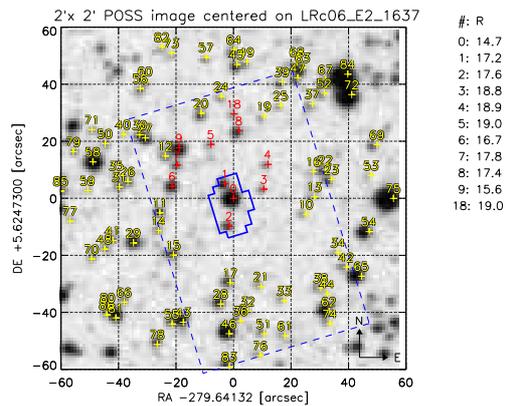}
  \end{center}
  \caption{Finding chart and contamination source map for CoRoT-33. Red numbers
           denote stars whose contribution to the observed flux was taken into 
	   account; yellow numbers denote stars whose contamination was checked 
	   but was found negligible. Star with number $0$ corresponds to CoRoT-33.
  }
  \label{fig:contamination}
\end{figure}

\begin{table}
\caption{IDs, coordinates, and magnitudes of CoRoT-33 from ExoDat (Deleuil et~al. 2009)}
\centering
\begin{tabular}{l@{\hskip2mm}c@{\hskip2mm}c}       
\hline\hline     
Designation     &  CoRoT-33 \\            
CoRoT window ID &  LRc06\_E2\_1637  & \\
CoRoT ID        &  105118236        & \\
2MASS           &  18383391+0537287 & \\
USNO-A2         &  0900-13338694 & \\ 
USNO-B1         &  0956-0378713  & \\ 
PPMXL           & 5484010357803959995 & \\
\\
\multicolumn{2}{l}{Coordinates} \\
\hline            
RA (J2000)  & 18$^\mathrm{h}$ 38$^\mathrm{m}$ 33.908$^\mathrm{s}$ \\
Dec (J2000) & +5$^\circ$ 37$^{\prime}$ 28.970$^{\prime \prime}$ \\
\\
\multicolumn{3}{l}{Magnitudes} \\
\hline
Filter & Mag \& Error & Source \\
\hline
$B$    & 15.705$\pm$0.587  & mean of several \\
       &                   & sources  \\
       &                   & (see ExoDat)\\
$R$    & 14.25 & USNO-A2.0 \\
$I$    & 13.5  & PPMXL \\
$J$    & 13.238$\pm$ 0.027 & 2MASS \\
$H$    & 12.811$\pm$ 0.026 & 2MASS \\
$K$    & 12.707$\pm$ 0.032 & 2MASS \\
\hline
\end{tabular}
\label{informationtable}      
\end{table}

%
\subsection{Radial velocity measurements}
\label{subsec:radial_velocities}

The radial velocity (RV) follow-up of CoRoT-33 was started with the HARPS spectrograph
\citep{mayor03} based on the 3.6-m ESO telescope (La Silla, Chile) as part of the ESO large program
188.C-0779. To monitor the Moon background light on the second fibre, HARPS was used with
the observing mode \emph{obj\_AB}, i.e., without simultaneous thorium-argon (ThAr). The  
exposure time varied between 25 minutes and 1 hour. A set of 8 spectra was recorded for CoRoT-33 
with HARPS between 28 June 2013 and 15 August 2013. We reduced the HARPS data and computed RVs with the
HARPS pipeline based on the cross-correlation technique \citep{baranne96,pepe02}. Radial velocities
were obtained by weighted cross-correlation with a numerical K5 mask which yielded the smallest
error bars. The  signal-to-noise ratio (S/N) per pixel at 5500{\,\AA} is in the range 2 to 7. 
The first and the last measurements were affected and corrected from  moonlight contamination.

Four additional RV measurements were acquired with the FIbre-fed \'Echelle Spectrograph
\citep[FIES;][]{Frandsen1999,Telting2014} mounted at the 2.56-m Nordic Optical Telescope (NOT) of
Roque de los Muchachos Observatory (La Palma, Spain). The observations were carried out on 2, 3, 4,
and 5 July 2014 under the CAT observing programme 79-NOT5/14A. We used the $1.3\,\arcsec$
\emph{med-res} fibre, which provides a resolving power of R\,$\approx$\,47,000 in the spectral range
3600\,--\,7400\,\AA. The weather was clear with seeing varying between 0.6 and 1.4\,$\arcsec$
throughout the whole observing run. We followed the observing strategy as described in
\citet{Buchhave2010} and \citet{Gandolfi2015} and took three consecutive exposures of 1200-1800 seconds per
epoch observation to remove cosmic ray hits. We also acquired long-exposed ($T_\mathrm{exp}= 22$-24~s)
ThAr spectra right before and after each epoch observation to trace the RV drift of the
instrument. The data were reduced using standard IRAF and IDL routines, which included bias
subtraction, flat fielding, order tracing and extraction, and wavelength calibration. The S/N
of the extracted spectra is 15--20 per pixel at 5500\,\AA. Radial velocity measurements were derived
via S/N-weighted, multiorder, cross-correlation with the RV standard star \object{HR\,5777}, and 
observed with the same instrument set-up as the target object.

The HARPS and FIES RV measurements are listed in Table~\ref{radialvelocity} 
along with their errors
and the Julian dates of the observations in barycentric dynamical time
\citep[BJD$_\mathrm{TDB}$, see][]{Eastman2010}.

%
\section{Analysis}
\label{subsec:analysis}

\subsection{Analysis of radial velocity data}

Full width half maximum (FWHM) and bisector span (BIS) of the 
cross-correlation function are also listed in Table~3 and show no 
significant variation in phase with the radial velocity. Furthermore the 
radial velocity peak-to-peak amplitude ($\sim$14 km/s) is larger than 
the FWHM excluding a blended or background binary scenario.

An offset between the two instruments was calculated as part of the orbit
fitting process and accounted for when combining the HARPS and FIES data sets.
We first tried fitting a circular orbit to the RV data. 
Free parameters were the offset value for each data set,  the true
$\gamma$-velocity and the amplitude ($K$) of the RV curve. 
The epoch and period were fixed to the  values of the 
photometric ephemeris (these were varied
within the errors). This fit yielded 
a considerably large reduced $\chi^2$-value: $\chi^2_\mathrm{min} = 166$.
We then also allowed the epoch and period to vary in the 
fitting procedure. 
The resulting fit yielded a slightly smaller  $\chi^2_\mathrm{min} = 122$. 
Interestingly, the period converged to a value that is compatible within $1\sigma$
uncertainties of the period obtained from the photometric transits, but the resulting
 epoch obtained from the RV-analysis differed by 194 minutes from the photometric value.
The error in the photometrically determined epoch is only 6 minutes, so the
epoch difference is significant at the 32$\sigma$ level. Assuming a circular orbit
the orbital fit to the RV data produced very large $\chi^2$-values or the RV data are
incompatible with the photometric ephemeris.

Therefore we tried fitting an eccentric orbit to the RV data. The results are quite satisfactory:
$\chi^2_\mathrm{min} = 1.76$. The orbit was found to have a slight eccentricity:  $e = 0.0700\pm0.0017$. 
The results of this fit can be found in Table~4.

Figure~3 shows the RV measurements -- after correcting for the RV offset 
and subtracting the systemic radial velocity -- phase folded to the 
orbital period (upper panel), along with the RV residuals (lower panel).

Zakamska et~al. (2011) found that a precise estimation of eccentricity requires 
that the radial velocity curve should have a S/N of 40, and the individual
RV measurements should have a precision better than 1\%. These
requirements are fulfilled in our analysis. Their Eq.~(7) gives an estimate of the
precision of the eccentricity determination, i.e., 
\begin{equation}
\log \sigma(e) = 0.48 - 0.89 \times \log (K \sqrt{N} / \sigma_\mathrm{obs})
\end{equation}
where $\sigma(e)$ is the expected precision on eccentricity, {$N=12$} is the number of the RV data 
points, $\sigma_\mathrm{obs}=0.046$ km/s is their average (median) uncertainty. Substituting our values, 
we get $\sigma(e)=0.011$ and thus our eccentricity determination and its error bar are 
reliable. The estimate of Zakamska et al. (2011) is the result of an average error estimate based only on RV, 
in our case the joint RV and light-curve fit yielded a smaller uncertainty range for the eccentricity
(Sect. 3.4). The relatively large $K/\sigma_\mathrm{obs}$ ratio enables us to achieve the 
aforementioned precision in the eccentricity.
      
\begin{table*}[th]
\caption{HARPS and FIES RV measurements of the CoRoT-33 system. 
         FWHM is the average full-width at half maximum of the 
         lines used (in Angstroms) and BIS is the corresponding bisector spans. 
         S/N is the signal-to-noise ratio of the individual spectra.
	 The FWHM of the FIES CCFs are wider than those extracted from the HARPS
         spectra because of the different resolutions of the two spectrographs.
	 }
\begin{tabular}{l@{\hskip2mm}c@{\hskip2mm}c@{\hskip2mm}cc@{\hskip2mm}cc@{\hskip2mm}cc@{\hskip2mm}c}
\hline
\hline
Instrument & BJD &  $v_\mathrm{rad}$ [km\,s$^{-1}$] & uncertainty [km\,s$^{-1}$] & FWHM & BIS & S/N \\
\hline
HARPS & 56471.70867  & 27.159 & 0.240 & 10.8 & -0.08 & 1.1 (moon)\\
HARPS & 56473.65935  & 14.793 & 0.033 & 10.1 & -0.04 & 7.4 \\
HARPS & 56474.75692  & 16.348 & 0.127 & 9.5  & -0.06 & 2.0 \\
HARPS & 56475.74925  & 23.725 & 0.051 & 9.9  & -0.07 & 5.1 \\
HARPS & 56509.71203  & 16.736 & 0.058 & 10.1 &  0.03 & 4.6 \\
HARPS & 56516.59952  & 24.455 & 0.040 & 9.9  & -0.12 & 6.0 \\
HARPS & 56518.67738  & 25.197 & 0.032 & 9.9  & -0.06 & 7.4 \\
HARPS & 56519.64904  & 18.248 & 0.085 & 10.0 &  0.04 & 3.3 (moon)\\
FIES  & 56841.58071  & 17.967 & 0.026 & 12.9 & 0.014 & 20 \\
FIES  & 56842.62444  & 25.419 & 0.033 & 12.7 & -0.056& 18 \\
FIES  & 56843.60496  & 28.120 & 0.036 & 12.5 & -0.044& 16 \\
FIES  & 56844.61337  & 24.856 & 0.046 & 12.6 & 0.006 & 15 \\
\hline
\end{tabular}
\label{radialvelocity}
\end{table*}

\begin{figure}
  \begin{center}
\includegraphics[width=\columnwidth,angle=0]{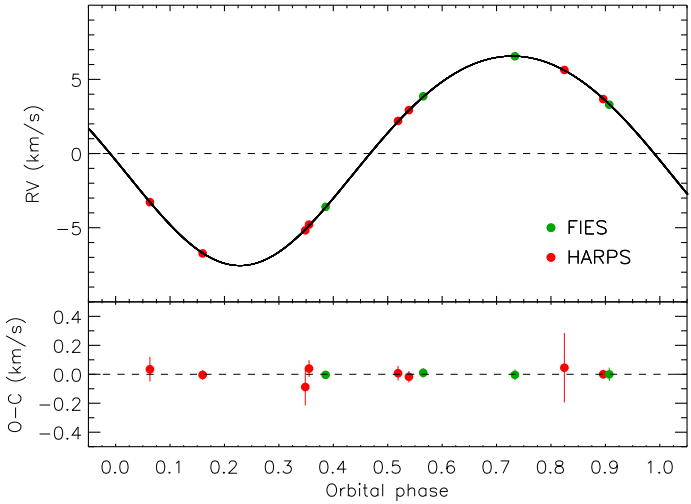}
  \end{center}
  \caption{Upper panel: phase-folded RV measurements of CoRoT-33. Red circles
           represent the HARPS-measurements while green circles are the data 
	   points obtained by FIES-instrument. Black solid line represents the
	   eccentric orbit fit. The RV points and the fit is shifted by the 
	   $\gamma$-velocity of the system. Lower panel: it shows the residuals 
	   of the fit. Vertical lines on the data points indicate their error bars.}
  \label{fig:rv_curve}
\end{figure}

\subsection{Spectral analysis of the host star}

We determine the spectroscopic parameters of the host star using the coadded HARPS and FIES
spectra. The S/N of the coadded data is relatively low, about 15 and 30 per pixel at 
5500{\,\AA} for HARPS and FIES, respectively, owing to the low S/N of each individual spectrum

We used the Spectroscopy Made Easy package (SME), version 4.1.2  (Valenti \& Piskunov 1996,
Valenti \& Fischer 2005) along with Atlas\,12 or MARCS\,2012 model 
atmospheres (Kurucz 2013, M\'esz\'aros
et~al. 2012) to determine the fundamental photospheric parameters iteratively. SME fits
the observed spectrum directly to the synthesized model spectrum and minimizes the
discrepancies using a nonlinear least-squares algorithm. The SME utilizes input from
the VALD database \citep{Piskunov1995,Kupka1999}. The uncertainties per measurement
using SME based on a sample
of more than 1000 stars was 
found by 
Valenti \& Fischer (2005) to be
44\,K for the effective temperature $T_\mathrm{eff}$, 0.06\,dex for the surface
gravity $\log g_*$, and 0.03\,dex for the metallicity $\mathrm{[M/H]}$. This is
in a statistical sense and somewhat optimistic. It was found from
Valenti \& Fischer (2005) 
that the scatter was systematically larger with an uncertainty
in $\log g_*$ of 0.1\,dex and a scatter that occasionally could reach 0.3\,dex. Fridlund
et~al. (2015, in preparation) have, in a systematic study of the CoRoT exoplanet
host stars, confirmed these numbers. Furthermore, the errors in $T_\mathrm{eff}$ are also
affected by the problems with determining the shape of the Balmer lines accurately
enough, as pointed out by Fuhrmann in a series of papers (see, e.g., Fuhrmann et~al. 
2011 and references therein; also Fridlund et~al. 2015, in preparation).

Torres et al. (2012) reported SME overestimates $\log g$ and correlates with $T_{\mathrm{eff}}$ and 
$\mathrm{[Fe/H]}$; see also Brewer et al. (2015) that this issue is eliminated. Therefore, we carried out an 
independent analysis of the spectrum using customized IDL software suite to fit the composite 
HARPS and FIES spectra with a grid
of theoretical model spectra from \citet{Castelli2004}, \citet{Coelho2005}, and
\citet{Gustafsson2008}, using spectral features that are sensitive to different photospheric
parameters. Briefly, we used the wings of the Balmer lines to estimate the effective temperature of
the star, and the Mg\,{\sc i} 5167, 5173, and 5184~\AA, the Ca\,{\sc i} 6162 and 6439 \AA, and the
Na \,{\sc i} D lines to determine $\log g_*$. The iron abundance $\mathrm{[Fe/H]}$ and microturbulent velocity
$v_{\mathrm{micro}}$ was derived by applying the method described in \citet{Blackwell1979} on
isolated Fe\,{\sc i} and Fe\,{\sc ii} lines. We adopted the calibration equations for Sun-like dwarf
stars from \citet{Doyle2014} to determine the macroturbulent velocity, $v_{\mathrm{macro}}$. The
projected rotational velocity $v \sin i_*$ was measured by fitting the profile of several clean and
unblended metal lines.

The two analyses provided consistent results well within the errors bars, regardless of the method 
and spectrum used. The final adopted values -- obtained as the weighted mean of the independent 
determinations -- are listed in Table~4. We found $T_{\rm eff}=5225\pm80$ K, $\log g_* = 4.4\pm0.1$ (cgs),  
$v \sin i_*=5.7\pm0.4$ km/s and a considerably high iron content:
${\rm [Fe/H]}=+0.44\pm0.10$. Using the \citet{Straizys1981} calibration scale for dwarf stars, the 
effective temperature of CoRoT-33 translates to a G9\,V spectral type.

%
\subsection{Estimation of stellar parameters}
\label{subsec:est_st_param}


The usual way to determine fundamental stellar parameters  is to combine
theoretical isochrones along with the measured  mean density of the star, 
stellar metallicity and effective surface temperature. The mean density can be
obtained from transit duration, the period and the impact parameter (e.g. Roberts
1899; Mochnacki 1981; Seager \& Mall\'en-Ornelas 2003; Winn 2010). 
Stellar models are then selected that are able to reproduce the
observed quantities. However, this method was found to be inadequate in our 
case because we have a grazing transit (see Sect.~3.4), and as a
consequence, a low-transit depth-to-noise ratio.

Therefore the effective temperature, metallicity and $\log g_*$ of the star 
were used 
to estimate the stellar mass and radius. 
We used the analytical stellar evolutionary tracks of Hurley et~al. (2000) 
which have about 2\% error relative to detailed numerical models, 
but are computationally fast. This
yielded $M_\mathrm{star} = 0.86\pm0.04M_\odot$ and $R_\mathrm{star} =
0.94^{+0.14}_{-0.08} R_\odot$ for the stellar mass and radius, respectively. The
age of the star is quite uncertain, but the object is definitely not young. This
is supported by the fact that we did not find the presence of lithium in the
spectrum. The lower age limit is 4.6~Gyrs and, most likely the star has an age 
of 11~Gyrs (see the derived stellar parameters in Table~4). 

As a sanity check we calculated the stellar radius independently of the stellar models.
If the stellar rotational axis is perpendicular to the orbital plane of the brown
dwarf, then $i_* = 85.5^\circ$ (cf. Sect.~3.4). Assuming that the
8.936 days modulation of the light curve is the stellar rotational period 
(cf. Sect 3.5), then the stellar radius can be computed from
\begin{equation}
v_\mathrm{rot} \sin i_* = \frac{2 \pi R_\mathrm{star} \sin i_*}{P_\mathrm{rot}}
\end{equation}
resulting in $R_\mathrm{star} = 1.06 \pm 0.09 R_\odot$. This is well within the
uncertainty range of isochrone-based value of 
$R_\mathrm{star} = 0.94^{+0.14}_{-0.08} R_\odot$.

%
\subsection{Transit light-curve analysis}
\label{subsec:transit_fit}

To measure the transit parameters, we removed the 
stellar variability from the light curve using the following procedure.
First, we applied a median filtering to replace the
outliers. We then fit the flux variations in a narrow vicinity of  the 
out-of-transit data (separately from transit to transit) with a parabola.
All points in and out of transit were divided 
by the corresponding parabola.
The width of the window which defined the `vicinity of the transit'
was found by visual inspection after assuming various trial lengths: $1D$,
$1.5D$, $2D$, etc.  up to $6D$ where $D$ is the transit duration. A $\pm2D$ 
window around each transit center was found to be the most appropriate.

The folded light curve using values binned by 201 s 
can be seen in Fig.~4. 

The modeling of the transit light curve is  challenging because of the 
faintness of the host star and small transit depth which result in a low
S/N (transit depth/average noise level is $\sim1$).
We used {\it Transit Light Curve Modeling Code} (TLCM, written by SzCs)
which utilizes the Mandel \& Agol (2002) model to describe the transit
light-curve shape. A genetic algorithm was used to optimize the fit (Geem et~al.
2001) and the error estimation was carried out using a simulated annealing 
chain of $10^5$ steps, starting from the best solution we found with the genetic 
algorithm procedure.

When scaled semi-major axis ($a/R_\mathrm{star}$) is treated as a free parameter, 
one gets the stellar mean density by rewriting Kepler's 3rd law, 
\begin{equation}
\rho_{\mathrm{star}} = \frac{3 \pi}{G P^2 (1+q)} \left( \frac{a}{R_s}\right)^3
\end{equation}
where $q=M_\mathrm{BD} / M_\mathrm{star}$ is the mass ratio. We find the star would 
have $\rho_{\mathrm{star}}\sim 29$ g\,cm$^{-3}$ which is incompatible with the
observed spectral type of G9V. The object G9V should have a mean density of about 1.27
g\,cm$^{-3}$. This fit also leads to 0.44 Jupiter-radii for the brown dwarf 
which results in a mean density of $\sim834$ g\,cm$^{-3}$. This is unrealistically 
high as no brown dwarf is known to have a mean density higher 200~g\,cm$^{-3}$ 
(cf. Table~1 and Ma \& Ge 2014).  
These results are independent of how limb darkening 
is treated. In our case
we used a quadratic limb darkening law fixed to  values 
based on Sing's (2010) tables. We also modeled the light curve allowing
the limb darkening  to vary as well as 
fixing one of the two limb darkening coefficients and allowing the 
other to be a free parameter.

The transit depth is 0.28\%, much shallower than the approximately 1\% for a
central  transit of a Jupiter-sized brown dwarf around a late G-type star.
The  contamination by other stars in the photometric aperture does not
explain the shallow transit depth. No additional contaminating source was 
revealed by the spectrum of the host star. The transit curve is V-shaped
and the  transit duration is just $\sim1.4$ hours, much shorter than the
expected $\sim3$  hours for a centrally transiting short-period substellar
object. This  suggests that we have a grazing eclipse and the transiting
object crosses the apparent stellar disk on a chord shorter than the
diameter of the star.

The transit duration is (cf. Seager \& Mall\'en-Ornelas 2003; Winn 2010):
\begin{equation}
D \approx \frac{P}{\pi} \times \frac{R_\mathrm{star}}{a} \times \sqrt{(1+k)^2 - b^2}
\sqrt{\frac{1-e^2}{1+e \sin \omega}}
\end{equation}
where $D$ is the measured transit duration ($D=1.4\pm0.1$ hours), $P$ is
the orbital  period, $e$ is the eccentricity, $\omega$ is the argument of
the periastron, $k$ is  the radius ratio and $b$ is the impact parameter.
All parameters are known from  photometry or from spectroscopy, except $k$
and $b$. Since the radius ratio must be positive we have the requirement
of $k>0$. Substituting the measured  values, we get that this requirement
is fulfilled only if $b>0.91$. This kind of a high impact parameter means we must
be dealing with a grazing transit scenario.

We conclude that the low S/N of the transit events prevents us from determining 
the transit shape with sufficient accuracy. We thus need to apply a penalty 
function (one may call a ``prior'') to force the light-curve solution
 to converge to the right values:
\begin{equation}
Q = \chi^2 + e^{{\frac{(\rho(a/R_\mathrm{star}) - 1.269)^2}{2 \times 0.210^2}}}
\end{equation}
and we minimized Q rather than $\chi^2$ of the light-curve fit. Here $\rho$
represents the mean density of the star calculated for every iteration step of the
fit via Eq. (3). The values 1.269 for the mean density and 0.210 for the width of the
distribution are the mean density of the star and its uncertainty (in g\,cm$^{-3}$),
derived from the stellar mass and radius obtained in Section~3.2.



In addition to the low S/N, the impact parameter is high, and it is difficult to fit 
the limb darkening coefficients if the impact parameter is 
larger than $\sim0.85$. Since 
the brown dwarf only crosses a fraction of the stellar disk, the transit
center is no longer flat-bottomed and the inner contact points disappear
(M\"uller et~al. 2013; Csizmadia et~al. 2013). 

Another difficulty is connected to the fact that our theoretical knowledge of limb 
darkening is very poor close to the limb of the star. For instance, based on 38 
Kepler light curves of transiting objects (12 of them are grazing), M\"uller et~al. 
(2013) established that the quadratic limb darkening coefficients have a larger 
disagreement from the theoretically predicted values of Claret \& Bloemen (2011) 
than the linear limb darkening coefficents.\footnote{Their limb darkening formula that we also use is 
$I = I_0 - u_1 (1-\mu) - u_2 (1-\mu)^2$, where $I_0$ is the intensity at the center 
of the apparent stellar disk, $u_1,~u_2$ are the linear and quadratic limb 
darkening coefficients, respectively; $\mu = \cos \gamma$; and $\gamma$ is the
angle between the line of sight and the surface normal vector at the stellar
surface point.} Difficulties in the theoretical limb darkening laws
occur mostly at the limb because of the quadratic term. 

This is further illustrated by the fact that Claret \& Bloemen (2011) and 
Sing (2010) used a plane-parallel stellar atmosphere model. When a more realistic
spherically symmetric model is used, Neilson \& Lester (2013) found a fast drop in 
the intensity close to the limb and the very edge of the star is predicted to be
much darker than in Claret \& Bloemen's model (for comparison of the plane-parallel 
and spherically symmetric models, see Figs. 2-5 of Neilson \& Lester 2013).
Since none of these new models have been checked against a large sample of eclipsing 
binaries and transiting objects covering a wide range of stellar temperatures, it is 
too early to accept these as a final description of the limb darkening law. In addition,
these theoretical models do not take the stellar spots 
and faculae into account which 
can modify the observable limb darkening coefficients (Csizmadia et~al.
2013). We should note that we can see evidence 
of the presence of spots on CoRoT-33, thus it is an active star.

Therefore we carried out two light-curve fits with the {\it Transit Light Curve 
Modeling Code} (TLCM). During the fits, our free parameters 
were the scaled semimajor axis, the impact parameter\footnote{We used $b  = \frac{a \cos i}{R_\mathrm{star}}
{\frac{1-e^2}{1+e \sin \omega}}$  to take the effect of eccentric orbit 
on the impact parameter into account.}, the radius ratio and the epoch. 
The limb darkening combinations $u_+ = u_a + u_b$ and $u_- = u_a - u_b$
were fitted once (cf. Csizmadia et~al. 2013 and Espinoza \& Jord\'an 2015), and then 
for a second calculation they were fixed at theoretical values of Sing (2010).
Eccentricity and argument of periastron could vary only within the limits of their 
uncertainties.

In the case of the adjusted limb darkening coefficients, the values of the coefficients 
are not constrained at all, but their uncertainties are large because of the low 
amplitude/noise ratio and the grazing transit. 
One cannot decide on a preferred solution 
based on the quality-parameter $Q$ alone. Taking into account that the fixed limb 
darkening coefficient solution contains two less free parameters, we have chosen that 
solution for subsequent discussion in the paper.

Then a joint fit of RV- and light-curve data by Exofast (Eastman et~al. 
2013) with adjusted limb darkening coefficients was carried out. This 
confirmed our solution in Table~4. The parameters of this brown dwarf 
should be refined in the future using better photometric measurements 
obtained with larger telescopes.

\begin{table}
\caption{Physical and geometrical parameters of the CoRoT-33 system. 
Inclination ($i$) was calculated from the $a/R_\mathrm{star}$ ratio and 
from the impact parameter $b$. The parameter $M^{1/3} / R$ can be 
calculated from the orbital period and from the $a/R_\mathrm{star}$ value 
(see e.g. Winn 2010).}
\begin{tabular}{l@{\hskip2mm}l}
\hline
\multicolumn{2}{l}{Determined from photometry} \\
\hline
Epoch of transit $T_{0}$ [BJD$-$2450000]  & 6676.3992$\pm$0.0037 \\
Epoch of periastron $T_{0}$ [BJD$-$2450000]&6677.7130$\pm$0.0140 \\
Orbital period (days)                     &  5.819143$\pm$0.000018\\
Duration of the transit (hours)           & 1.4 hours\\
Depth of the transit (\%)                 & 0.28\% \\
\hline
\multicolumn{2}{l}{Determined from RV measurements} \\
\hline
Orbital eccentricity $e$  & $0.0700\pm0.0016$ \\
Argument of periastron $\omega$ [deg] & $179.3\pm0.87$ \\ 
RV semi-amplitude $K$ [\kms] & 7.0609 $\pm$ 0.0094 \\
Systemic velocity  $V_{\gamma}$ [\kms], HARPS &  21.5375 $\pm$ 0.0158  \\
Systemic velocity  $V_{\gamma}$ [\kms], FIES  &  21.5544 $\pm$ 0.0222 \\
O$-$C residuals$^a$ [\ms] & 36 \\
\hline
\multicolumn{2}{l}{Determined from spectral analysis of the star} \\
\hline
$T_\mathrm{eff}$  [K]        & 5225$\pm$80   \\
$\log g_*$ [cgs] & 4.4$\pm$0.1   \\
$\mathrm{[Fe/H]}$            & 0.44$\pm$0.1  \\
$\mathrm{[Ni/H]}$            & 0.4$\pm$0.1   \\
$\mathrm{[V/H]}$             & 0.4$\pm$0.1   \\
$\mathrm{[Mg/H]}$            & 0.4$\pm$0.1   \\
$\mathrm{[Ca/H]}$            & 0.3$\pm$0.1   \\
$\mathrm{[Si/H]}$            & 0.2$\pm$0.1   \\
$v \sin i_*$ [km\,s$^{-1}$]    & 5.7$\pm$0.4   \\
Spectral type                & G9V           \\
$V_\mathrm{mic}$             & 0.86$\pm$0.1 km/s \\
$V_\mathrm{mac}^b$             & 2.7$\pm$0.6 km/s\\
\hline
\multicolumn{2}{l}{Determined from light curve modeling} \\
\hline
$a/R_\mathrm{star}$ & 13.23$\pm$1.17\\
$b$                 & 1.04$\pm$0.06\\
$i_{planet}$ [deg]  & 85.5$\pm$0.5\\
$k$                 & 0.12$\pm$0.04\\

contamination [\%]  & 13$\pm$4$^c$ \\
\hline
\multicolumn{2}{l}{Combined results} \\
\hline
Stellar mass $M_\mathrm{st}$ [solar]            & 0.86$^{+0.04}_{-0.04}$ \\
Stellar radius $R_\mathrm{st}$ [solar]          & 0.94$^{+0.14}_{-0.08}$ \\
Stellar age [Gyrs]                       & $>4.6$         \\
Orbital semi-major axis $a$ [AU]         & 0.0579$^d$ \\
Brown dwarf mass   $M_\mathrm{BD}$ [M$_J$ ]     & 59.0$^{+1.8}_{-1.7}$  \\
Brown dwarf radius $R_\mathrm{BD}$ [R$_J$ ]     & 1.10 $\pm$ 0.53 \\
Brown dwarf mean density $\rho_\mathrm{BD}$ [g\,cm$^{-3}$] &  $55\pm29$\\
\hline
\hline
\end{tabular}
\tablefoot{\tablefoottext{a}{Root mean square of the residuals of the 
RV curve.} \tablefoottext{b}{Fixed at this value, based on the calibration by
Doyle et~al. (2014).} \tablefoottext{c} {Calculated from period and masses via 
Kepler's third law, not from RV.} \tablefoottext{d} {From modeling results.}}
\label{planparams}
\end{table}

\begin{table*}
\caption{Results of three light-curve solutions of CoRoT-33. {\it ldc} stands for
limb darkening coefficients. Inclination is calculated from the scaled semi-major
axis, impact parameter, eccentricity and argument of periastron.}
\begin{tabular}{llll}
\hline
\hline
Parameter           & fixed ldc (TLCM)       & adjusted ldc (TLCM) &  ExoFast\\
\hline
$a/R_\mathrm{star}$ & 13.23$\pm$1.17  & 13.24$\pm$0.33 &  14.20$\pm$1.25  \\
$b$                 & 1.04$\pm$0.06   &  1.00$\pm$0.05 &  1.053$\pm$0.065 \\
$i_{planet}$ [deg]  & 85.5$\pm$0.5    & 85.7$\pm$0.4   &  85.70$\pm$0.53  \\
$k$                 & 0.12$\pm$0.04   & 0.078$\pm$0.034&  0.139$\pm$0.039 \\
$u_+$               & 0.57            & 0.13$\pm$0.36  &  -\\
$u_-$               & 0.12            & 0.33$\pm$0.99  &  -\\
$u_1$               &                 &                &  0.538$\pm$0.053  \\
$u_2$               &                 &                &  0.171$\pm$0.051  \\
contamination [\%]  & 13$\pm$4        & 13$\pm$4       & 13\% (fixed) \\
Q                   & 1.257           & 1.257          & \\
\hline
\end{tabular}
\label{limbdarkenoingissue}
\end{table*}

The fit is shown in Table~4 and Fig.~4.

%
\subsection{Stellar variability analysis}
\label{subsec:variability}

The light curve of CoRoT-33 exhibits periodic and quasiperiodic 
flux variations with a peak-to-peak amplitude of about 3\% (Fig.~1, 
upper panel). The Lomb-Scargle periodogram of the light curve is 
shown in Fig.~1. The strongest peak occurs at $\nu$ = 0.1117 d$^{-1}$ 
($P$ = 8.95 d). Toward higher frequencies, there are two additional 
significant peaks at $\nu$ = 0.2248 d$^{-1}$ ($P$ = 4.44 d) and 
$\nu$ = 0.3337 d$^{-1}$ ($P$ = 3.00 d). The dominant peak is 
consistent with the expected rotational period, $P_\mathrm{rot}$ of the 
star calculated using the spectroscopically measured rotational 
velocity and stellar radius. Therefore, this signal is most likely 
due to spots since CoRoT-33 should have a modest level of magnetic 
activity given its relatively fast rotation rate. Interestingly, 
this rotational period is $\sim 3/2 \times P_\mathrm{orb}$. The 
other two peaks correspond to the first and second rotational 
harmonics ($P_\mathrm{rot}$/2, $P_\mathrm{rot}$/3, respectively). 
This probably reflects a complex pattern on the stellar surface.
For instance, $P_\mathrm{rot}$/2 can result from two spot groups 
on opposite sides of the star.


Following the guidelines described in \citet{McQuillan2013} and 
revised in \citet{McQuillan2014}, we use the autocorrelation 
function (ACF) to confirm these results. The ACF of the 
CoRoT-33 light curve 
shows a strong correlation peak at the rotational period of approximately nine days,
consistent with the Lomb-Scargle periodogram, 
followed by a sequence of additional local maxima at 
integer multiples of this value. This is the result of the 
viewing geometry which repeats after multiple rotations (Fig.~1, lower 
panel). We estimate the position of the peaks by fitting Gaussian 
functions to the sequence of ACF maxima, and define the rotation 
period as the slope of a straight-line fit to the ACF peak positions 
as a function of peak numbers. This yields a stellar rotation period of 
$P_\mathrm{rot}=8.936\pm0.015$\,days.

\begin{figure}
  \begin{center}
\includegraphics[width=9.62cm,angle=0]{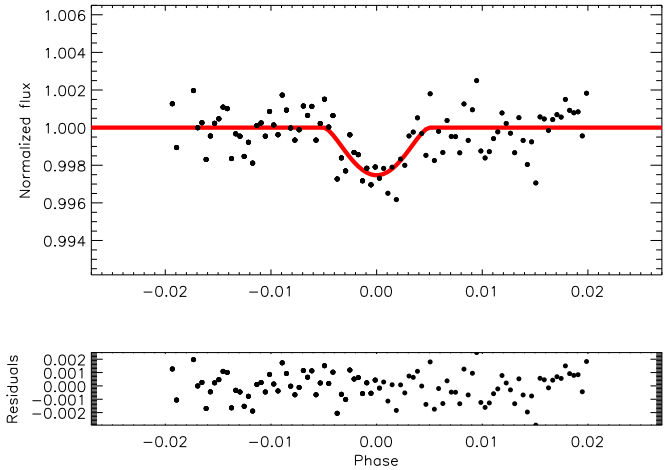}
  \end{center}
  \caption{Light-curve solution (solid red line) and phase-folded, binned data points (black circles) 
           of CoRoT-33b transits. Lower panel shows the residuals of the fit. Note that transit depth is
	   only 0.28\% on an $R=14.25$ magnitude star.}
  \label{fig:Light_curve_solution}
\end{figure}

We also investigate the periodicities in the light curve using the 
package {\sl MUFRAN\/} developed by Koll\'ath (1990). This software 
is an efficient tool for detecting periodic patterns in time series and is 
based on the Fourier transform.

\begin{figure}
  \begin{center}
\includegraphics[width=7.62cm]{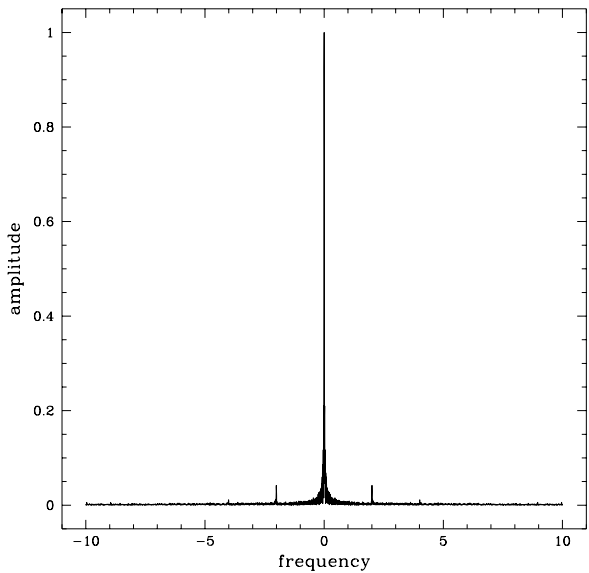}
  \end{center}
  \caption{Window function of CoRoT-33.
  }
    \label{sp_window}
\end{figure}

Owing to temporal gaps in the data some false frequencies (aliases) can 
appear in the Fourier power spectrum. These alias frequencies are 
centered on the real signals offset from those peaks as indicated by 
the spectral window function of the Fourier transform. The spectral 
window (Fig.~\ref{sp_window}) indicates a negligible aliasing caused 
by the data sampling.

\begin{figure}
  \begin{center}
\includegraphics[width=7.62cm]{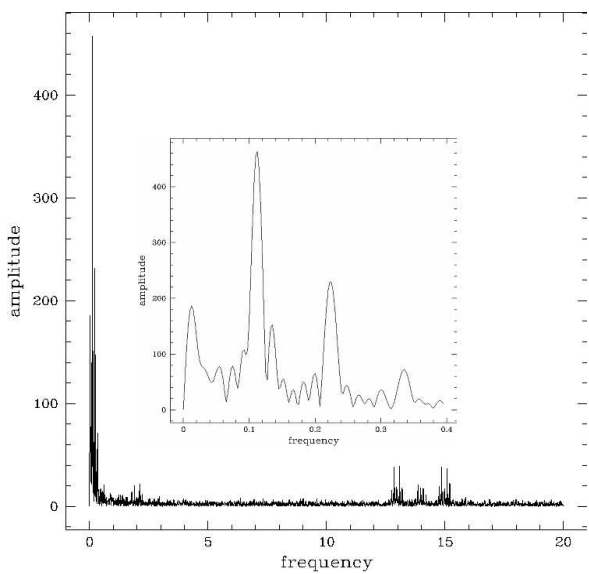}
  \end{center}
  \caption{Fourier amplitude spectrum of the light curve of CoRoT-33. 
  Frequencies are in cycle/day units. The insert shows a zoom into 
  the low-frequency part.
  }
  \label{fourier_amp}
\end{figure}

The Fourier amplitude spectrum of the whole data set is shown in 
Fig.~\ref{fourier_amp} for this wide interval. The multiple peaks 
near 14~c/d are artifacts due to scattered earthshine arising from
the 103 min orbit of the  CoRoT satellite.
For better visibility, the low-frequency part of 
the amplitude spectrum is shown in the insert of
Fig.~\ref{fourier_amp}. This Fourier analysis confirmed the 
8.936 day rotational period and the presence of its harmonics. The 
period analysis was carried out separately for the first and 
second half of the data set, too, and it showed that the main 
frequencies do not change with time.

%
\subsection{Search for beaming effect}
\label{subsec:beaming}

One can ask whether one of the frequencies in the Fourier spectra is 
related to beaming effect (Zucker et~al. 2007). Mazeh \& Faigler (2010) 
cleaned and binned the light curve to 100 points to see this effect in 
CoRoT-3. Unfortunately, we cannot follow their approach. 
We can  estimate the magnitude of this effect using their formula:
\begin{equation}
F(t) = F_0 + A_{\rm b} sin(\omega (t - t_0)) - A_{\rm r} cos(\omega (t - t_0)) - A_{\rm e} cos(2\omega (t - t_0))
\end{equation}
where $\omega = 2 \pi / P_\mathrm{orbital}$, and $A_{\rm b}$, $A_{\rm r}$, $A_{\rm e}$ are 
the amplitudes of the beaming effect, the variation coming from reflected light
and from the ellipsoidal shape of the star. All these effects act on the same
time-scale, therefore we see a periodic signal at the orbital period and its
half value. Using the formulae of Mazeh \& Faigler (2010), we expect $A_{\rm b} = 94$~ppm,
$A_{\rm r} = 9$~ppm, and $A_{\rm e} = 54$~ppm for a total peak-to-peak amplitude 
of 231~ppm. Since the
eccentricity is small, we can neglect its effect. Notice that these 
estimates are uncertain by several percent (Mazeh \& Faigler 2010).

For an albedo of 0.1 and 0.9, the equilibrium temperature of the 
brown dwarf object would be between 1400-800~K in this system. 
Baraffe et~al. (2003) predicts a surface temperature of 1100 -- 1200 K for a 
brown dwarf with a  mass and age consistent with our object. Thus the brown dwarf contributes 
only 0.2 ppm to the observed visible light so it
can be neglected.  

Using the approximate formula of Aigrain et~al. (2009), we get that 
the two hours average noise level of CoRoT at the magnitude of 
CoRoT-33 is about 365~ppm (not taking the aging of the 
CCD detector into account). This means that we have to apply very strong binning to 
see the effect. To detect this signal at a S/N of 3  we had to apply 
a binning of two days. Therefore we choose 
a different approach from that of Mazeh \& Faigler (2010);  
namely we utilize the Fourier-spectrum.


The cleaned light curve was taken again and a five-point median-filtering was carried out.
We discovered a significant peak at $5.93\pm0.32$ days with an amplitude 
of 318~ppm in the Fourier-spectrum after a cleaning process. This peak is
significant at the 13$\sigma$ level (its significance was  estimated by
using Eq.~(21) of Kjeldsen \& Frandsen 1992). Its amplitude is 50\%
bigger than our expectation for the peak at the orbital period, but the
expectation is also uncertain by several percent. Although this peak is
quite close to the orbital period, it can also be identified as a
harmonic of the rotational period of the star; the period ratio of this
peak and the stellar  rotational period is exactly 1.5. Further study is
required to separate the stellar  rotational modulation and the beaming
effects.

%
\subsection{A search for occultations}
\label{subsec:occultation}

A brown dwarf in a close-in orbit may be occulted by the parent star
(e.g. Winn 2010) which could be 
helpful to constrain the impact parameter and other properties of the 
system with higher accuracy. We searched for these kinds of occultations of the 
brown dwarf in the light curve of CoRoT-33 using a Bayesian model 
selection (e.g. Kass \& Raftery 1995, Gregory 2010), wich are similar to the 
approaches used in Parviainen et~al. (2013, 2014) and Gandolfi et~al. 
(2015). The stellar variability was taken into account using Gaussian 
processes (Roberts et al. 2013; Gibson et~al. 2011; Rasmussen 
et~al. 2006). The nonzero orbital eccentricity means that the impact 
parameter for the occultation, $b_\mathrm{o}$, can be significantly 
different than the impact parameter for the transit, $b_\mathrm{t}$, or 
there might be no occultation. The technical details of this search are 
given in the Appendix. Our analysis was unable to find the occultation 
signal. The missing occultation signal can also be 
consistent with Baraffe et al.'s (2003) models.

%
\section{Tides and stellar rotational properties}
\label{sec:rotation}

The rotation period is small for a G9V star of its age. The braking 
associated with the activity of a G9V star is not as efficient as it 
should be if we take the single-star scenario of Bouvier et~al. (1997)
or Mamajek \& Hillenbrand (2008). 
Because of the heavy companion orbiting the star at a short 
distance, the tidal torques on the star are strong enough to         
accelerate the stellar rotation. Several factors concur to make this 
system one of the best suitable for the study of the interplay of 
magnetic braking and tidal evolution: the high mass of CoRoT-33b, 
the short distance from it to the star, and the age of the system. 

The simulations done by Ferraz-Mello et~al. (2015) show that in a 
system like CoRoT-33, the magnetic braking may be very efficient in 
the beginning and this drives the rotational period to a value somewhat 
larger than the current observed stellar rotational period, when braking and tidal 
evolution equilibrate themselves. Subsequently, the system evolves 
losing energy, and the companion orbit spirals down toward the star 
and both the orbital period of the companion and the rotation period 
of the star slowly decrease for the remaining life of the 
system, and it can possibly reach such commensurability just by chance.
 The fact that the orbital period and the stellar rotation 
period are strikingly close to the 2/3 ratio is not predicted in 
usual spin-orbit dynamical theories. B\'eky et~al. (2014) listed six 
systems with a hot Jupiter where the planet and the star exhibit 
similar synchronization between these periods. They consider that the 
stellar differential rotation profile may happen to include a period 
at some latitude that is commensurable to the planetary orbit.
These influence the displacement of the star features responsible 
for the periodic variation of the light of the star. The quality of 
our characterization of CoRoT-33 is in favors of it playing a key role 
in the study of these sorts of interactions and their influence on the 
tidal evolution of the system.

%
\section{Conclusions}
\label{subsec:conclusions}

\begin{figure}
  \begin{center}
\includegraphics[width=7.62cm]{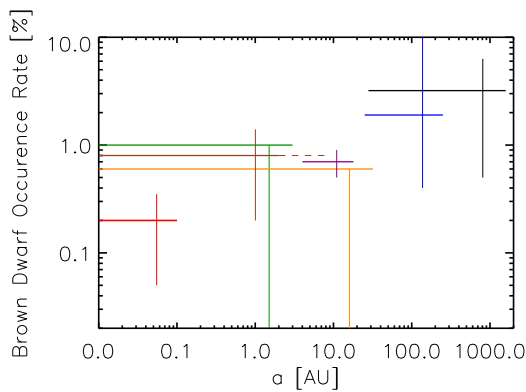}
  \end{center}
  \caption{Occurrence rate of brown dwarfs as companions to solar-like stars vs orbital separation ($a$).
           Horizontal lines denote the found occurrence rates and the orbital 
	   separation range investigated while vertical lines denote the error bars. Different colors mark  
	   different studies; black is Metchev \& Hillenbrand (2009); blue is Lafreni\'ere et al. (2007), 
	   Grether \& Lineweaver (2006), orange is Sahlmann et al. (2011), brown is Wittenmyer et al. 
	   (2009), violet is Patel et al. (2007), red is this study. These rather rough estimates indicate 
	   a tendency: the occurrence rate increases with increasing orbital separations.
  }
  \label{fig:contamination}
\end{figure}

We report the detection of a transiting $59.0^{+1.8}_{-1.7}$ Jupiter-mass brown 
dwarf that orbits a G9V star in $\sim$5.82 days. This is the 10th 
transiting object in the so-called brown dwarf desert not counting the 
transiting double brown dwarf system 2M0535-05 here). This desert is 
starting to be populated by more and more objects in recent years. The 
radius of the brown dwarf is $1.10\pm0.53$ Jupiter radii and its mean 
density is $55\pm29$ g\,cm$^{-3}$. The object is close to the 
65~$M_\mathrm{Jup}$ limit which separates the lower mass, 
deuterium-burning brown dwarfs from the higher mass, lithium-burning 
brown dwarfs. The host star seems to be an evolved, old, metal-rich star. The 
radius of the brown dwarf is not known to high precision because of a 
grazing transit. On the other hand, the mass is measured with higher 
accuracy from the RV observations, and its mass uncertainty is dominated by 
the uncertainty in the stellar mass. Better photometric precision using 
larger telescopes is required to improve the radius and thus 
density determination.

The statistical analysis of 65 brown dwarf companions to stars with 
orbital period less than 100 days by Ma \& Ge (2014) showed that 
their host stars are not metal rich stars. This may be related 
to a different formation scenario of brown dwarfs compared to giant 
planets. The host stars of giant planets seem to be more metal rich 
(Johnson et~al. 2010). However, CoRoT-33 is metal rich with 
$\mathrm{[Fe/H]}=+0.44\pm0.10$ and its abundance may exceed the 
previous record holder, HAT-P-13 with $\mathrm{[Fe/H]}=0.41 \pm 
0.08$ (Bakos et~al. 2009; Ma \& Ge 2014)\footnote{The brown dwarf 
HAT-P-13c is not transiting, it was detected by RV
 (Bakos et~al. 2009).}.

The light curve of CoRoT-33 shows a rotational modulation with a 
period of 8.936 d with a peak-to-peak amplitude of $\sim$3\%. This 
measured rotational period agrees with the calculated value 
based on the $v\sin i_*$ measurement and its stellar radius from 
isochrones (assuming $i_* \approx i_{BD}$). The Fourier spectrum is complex,
 making the target ideal 
for future studies of spot activity on this star.  In particular, 
multicolor photometry using large telescopes
can help to  determine 
the spot temperatures. 

The rotational period of the star is very close to the 2:3 
commensurability with the orbital period of the companion brown dwarf 
and this system is an interesting test case for checking and calibrating 
theories of the physical interactions between one star and a close-in 
companion. This relatively old system shows an eccentric orbit; the 
study of the tidal evolution of this system (Ferraz-Mello et~al. 2015) 
shows that because of the distance between the brown dwarf and the star 
(0.0626 AU), an existing initial eccentricity is not damped to zero 
during the stellar lifetime, however, the presently observed rotational 
period of the star cannot be explained without taking the interplay 
between magnetic braking of the star and tidal forces into account.

One can assume that close-in hot Jupiters and brown dwarfs have the 
same transit detection bias since they both have approximately the 
same radius. Therefore, the ratio of the number of these detected 
systems should reflect the respective number frequencies of these objects.  
Until now, three brown dwarfs (CoRoT-3b, -15b and 33b) have been
detected from the CoRoT data, and 21 hot and normal Jupiters 
(CoRoT-1b, -2b, -4b, -5b, -6b, -9 -- -14b, 16b -- -21b, -23b, 
25b -- -29b) most with periods $P<10$ days.\footnote{CoRoT-9b and 
-10b have an orbital period of 95.23 days and 13.24 days, respectively.} 
The relative frequency, based on the CoRoT-sample and hence the observational 
biases are removed, is of $\approx$ 14\%, but one has to take into 
account that so far we have only a small sample statistic.

If we take all transit surveys, then the brown dwarf/hot Jupiter  ratio
falls down to 0.05\%. For instance, the WASP survey has detected one brown
dwarf (see Table~1) and 96 hot Jupiters. HAT-P-survey has detected 33 hot
Jupiters and none with a mass exceeding 8~$M_\mathrm{Jup}$; there is no 
transiting brown dwarf in the HAT-P-sample (see also footnote 4). It is
not well  known how the targets for follow-up and the stellar samples are 
selected in these surveys. Better statistics should come from Kepler  once
the follow-up measurements for the transit candidates are  completed which
may resolve the discrepancy between ground-based  and CoRoT observations.
A major factor can be that ground-based  surveys are biased for the
detection of transits smaller than 1\%.

The true occurrence rate of hot Jupiters is $1.2\pm0.4$\% around FGK 
dwarf stars for periods $P<10$ days (Wright et~al. 2012). Thus, the 
relative frequency of brown dwarfs to hot Jupiters can be scaled to 
the true frequency which means that the actual occurrence rate of brown 
dwarfs as companions to FGK dwarfs would be $\sim0.2$\% in the short 
period ($P<10$ days) range. This is much smaller than the frequency 
rate of brown dwarfs as FGK companions at larger distance (see Introduction).
One can plot the different brown dwarf occurrence rate estimates mentioned
in the Introduction and determined here as a function of the star-companion 
distance. The result is shown in Fig.~7. A fit to those data yielded that the 
brown dwarf occurrence rate ($f$) around FGK dwarfs can be characterized roughly as 
$f = 0.55^{+0.8}_{-0.55} (a/1\mathrm{AU})^{0.23\pm0.06}$, where $a$ is the semimajor
axis of the orbit. This result is not robust yet because the occurrence rate
estimates suffer from small number statistics, but a tendency might be visible
already. More observational studies are needed to establish the occurrence rate-orbital separation
relationship for solar-like star$+$brown dwarf pairs.

The early indication is that the occurrence rate of close-in brown dwarf
($P<10$ days) is six times  smaller than that of hot Jupiters in the same
period range. It is not clear  whether this relative occurrence rate is a
consequence of the primordial  conditions (i.e., much fewer brown dwarfs
were formed in the protoplanetary disk than giant planets), or is caused by 
the higher efficiency of engulfing the companion by the host stars due to
fast spiralization, as Armitage \& Bonnell (2002) presumed. It is also possible 
that observational biases act: for instance, CoRoT's stellar sample or its 
follow-up strategy or its higher photometric precision than ground-based surveys 
were simply more sensitive to brown dwarfs.

For larger sized planets, the planet-frequency decreases for shorter periods 
(Dong \& Zhu 2013). The brown dwarfs seems to follow the same pattern, too, 
but we think there is an indication that the frequency of close-in brown dwarfs
 drops more steeply than that of close-in giant planets. However, further detection
studies are needed to establish the true frequency of close-in brown
dwarfs.

%
\begin{acknowledgements}
The team at IAC acknowledges support by grant AYA2012-39346-C02-02 of 
the Spanish Secretary of State for R\&D\&i (MICINN). R.A. acknowledges 
the Spanish Ministry of Economy and Competitiveness (MINECO) for the financial 
support under the Ram\'on y Cajal program RYC-2010-06519. This research 
has made use of the ExoDat database, operated at
LAM-OAMP, Marseille, France, on behalf of the CoRoT/Exoplanet
program. 
This publication makes use of data products from the Two Micron All
Sky Survey, which is a joint project of the University of
Massachusetts and the Infrared Processing and Analysis
Center/California Institute of Technology, funded by the National
Aeronautics and Space Administration and the National Science
Foundation.
This research has made use of NASA's Astrophysics Data System.
L.~Szabados was supported by the ESTEC Contract No.\,4000106398/12/NL/KML.
The first author thanks the Hungarian OTKA Grant K113117. 
The German CoRoT Team (TLS and the University of Cologne) acknowledges DLR 
grants 50 OW 204, 50 OW 0603 and 50QP07011. 
This research has benefited from the M, L, T, and Y dwarf compendium housed 
at DwarfArchives.org.
HP has received support from the Leverhulme Research Project grant RPG-2012-661.
A.S. is supported by the European Union under a Marie Curie
Intra-European Fellowship for Career Development with reference
FP7-PEOPLE-2013-IEF, number 627202 and by Funda\c c\~ao para a Ci\^encia
e a Tecnologia (FCT) through the research grant UID/FIS/04434/2013. 
JMA  acknowledges support by CNES grant 251091. The prompt 
and valuable report of an anonymous referee is acknowledged.


\end{acknowledgements}

%


\begin{appendix}

\section{Bayesian model selection}

Here we describe our methodology which was used for searching the 
possible occultation of the brown dwarf.

We consider two models, $M_0$ without an occultation signal and $M_1$ with an 
occultation signal. We also use the Bayes factor $B_{10}$, the ratio of the Bayesian 
evidences $Z_M$, to assess whether the occultation model is significantly preferred 
over the non-occultation model. The evidence $Z_M$ for a model $M$ is calculated by 
marginalizing the posterior probability over the model parameter space,
\begin{equation}
 Z_M = \int P(\pvec|\data)\,\ud\pvec = \int P(\pvec) P(\data|\pvec)\,\ud\pvec,
\end{equation}
where \pvec is the model-specific parameter vector, $D$ is the observational data, 
$P(\pvec|D)$ is the posterior probability density, $P(\pvec)$ is the prior density, 
and $P(D|\pvec)$ is the likelihood for the observational data.

We assume the photometric noise to be normally distributed, and express the likelihood as
\begin{equation}
  P(\data|\pvec)=-\frac{1}{2} \left( n_\mathrm{D} \ln 2\pi +\ln\det\covmat + \vec{r}^\mathrm{T} \covmat^{-1} \vec{r}\right),
\end{equation}
where $n_\mathrm{D}$ is the number of data points, \covmat is the covariance matrix, 
and $\vec{r}$ is the observed-measured residual vector. The covariance matrix elements are 
defined by the Gaussian process (GP) covariance kernel, $k(\vec{x_i},\vec{x_j},\vec{\phi})$, where $x$ are 
input parameter vectors for each data point, and $\vec{\phi}$ is a 
kernel hyperparameter vector. 

We only use one GP input parameter, mid-exposure time. We decided not to marginalize over 
the GP hyperparameters (we fix them to values optimized to the light-curve), but we did 
decide to use two different GP kernels to assess the sensitivity of our analysis 
on the choice of kernel. The two kernels used were the squared exponential (SE) kernel
\begin{equation}
 k_\mathrm{SE}(t_i,t_j, h, \lambda) = h^2 \exp\left(-\frac{(t_j-t_i)^2}{\lambda}\right),
\end{equation}
and the exponential (E) kernel
\begin{equation}
 k_\mathrm{E}(t_i,t_j, h, \lambda) = h^2 \exp\left(-\frac{|t_j-t_i|}{\lambda}\right),
\end{equation}
where $t$ is the mid-exposure time, $h$ is the output scale, and $\lambda$ is the input scale. 
The SE kernel leads to infinitely-differentiable smooth functions, while the E kernel leads to 
once-differentiable functions allowing for sharper changes (which is a more realistic choice 
considering the noise properties of CoRoT light curves).

\subsubsection{Parameterization and priors}
We construct the priors for the orbital parameters and 
radius ratio based on the corresponding marginal posteriors from the transit and RV modeling, 
which leaves the surface flux ratio as the only truly unconstrained parameter.

In parallel with the main model selection analysis, we carry out a more explorative occultation 
search by mapping the $B_{10}$-space as a function of sliding prior on $\omega$. 
This corresponds roughly to making $n$ model comparisons for a set of propositions, each 
differing in the prior set on $\omega$. We use an uniform prior defined by its center and 
width, and let the prior center slide from 0~to~$2\pi$. 


\section{Results}
We did not find support for the occultation model over the no-occultation model when we constrain $\omega$ using a 
prior based on RV and transit modelling. The geometry, $\omega \sim 180^\circ$, combined with the high impact 
parameter makes the detection of even a very strong signal unlikely.

\begin{figure}
 \centering
 \includegraphics[width=\columnwidth]{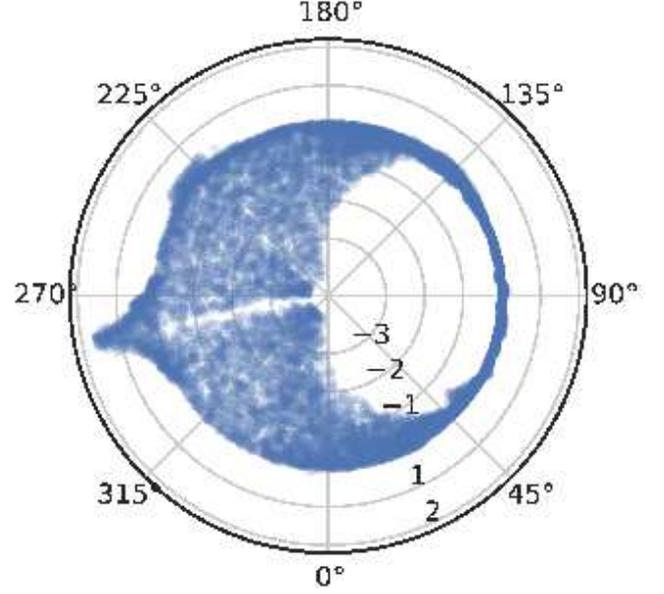}
 \caption{Differences between the M$_0$ and M$_1$ log posterior samples with the SE kernel 
  (corresponding to the log posterior ratios) for a set of 15000 posterior samples mapped as a function of 
  argument of periastron (angle). A uniform prior from 0~to~$2\pi$ has been set on $\omega$, and all the 
  parameters have been drawn from their corresponding priors. The shown value, sample log posterior 
  difference, does not correspond to the Bayesian evidence $Z$, but is used as an explorative tool. The 
  radial spread in the difference between $\omega$ values of $\pi$ and $2\pi$ is explained by the orbital 
  geometry. The occultation signals are stronger within this interval than on the other half, and the data 
  is able to exclude these signals leading to small log posterior ratio. The occultation signals between 
  $\omega$ of 0 and $\pi$ are too faint to be excluded by the data, and the with-occultation corresponds to 
  the occultation model within uncertainties.}
 \label{fig:eclipse_search}
\end{figure}

The mapping of $B_{10}$ as a function of $\omega$ results in a tentative occultation signal candidate near $270^\circ$, as 
shown in Fig.~\ref{fig:eclipse_search}. The result is curious, since this is the only geometry where an occultation could 
in theory be detected, while a false signal of instrumental or astrophysical origin could present itself anywhere in 
$\omega$-space. However, the small number of orbits covered combined with the faintness of the signal candidate makes this 
impossible to confirm. If the occultation signal were to be real, it would be discrepant with the RV observations. This 
could be alleviated if the RV observations were to contain an unknown noise source, but the increase in the uncertainty 
per RV observation would need to be around 0.2-0.3~km/s.

The two GP kernels yield very similar results, meaning our analysis is not sensitive to the choice of the GP kernel.

\end{appendix}

\end{document}